\documentclass[tightenlines,superscriptaddress,eqsecnum,floats,nofootinbib,showpacs]
{revtex4}
\usepackage{amssymb}
\usepackage{stmaryrd}
\usepackage{amsmath}
\usepackage{amsfonts}
\usepackage{mathrsfs}

\usepackage{amsmath,amssymb,amsfonts}
\usepackage{graphicx}
\usepackage{ulem}
\usepackage{color} % for colored text
\usepackage{setspace}
\def\be{\begin{equation}}
\def\ee{\end{equation}}
\def\ba{\begin{eqnarray}}
\def\ea{\end{eqnarray}}

\newcommand{\figref}[1]{Fig.~\ref{#1}}

\begin{document}

%\date\today
%\preprint{????}

\title{Thermodynamics of isolated horizons in loop quantum gravity}

\author{Shupeng Song}
\thanks{songsp@bit.edu.cn}
\affiliation{School of Physics, Beijing Institute of Technology, Beijing 100081, China}
\author{Gaoping Long}
\thanks{gaopinglong@mail.bnu.edu.cn}
\affiliation{Department of Physics, South China University of
Technology, Guangzhou 510641, China}
\author{Cong Zhang}
\thanks{czhang@fuw.edu.pl}
\affiliation{Faculty of Physics, University of Warsaw, Pasteura 5, 02-093 Warsaw, Poland}

\author{Xiangdong Zhang\footnote{Corresponding author. scxdzhang@scut.edu.cn}}
\affiliation{Department of Physics, South China University of
Technology, Guangzhou 510641, China}

\begin{abstract}
The statistical mechanical calculation of the thermodynamical properties of nonrotating isolated horizons are studied in the loop quantum gravity framework. By employing the Hawking temperature and horizon mass of isolated horizons as physical inputs, the microcanonical ensemble associated with the system are well established. As a result, the black hole entropy
and other thermodynamical quantities can be computed and are consistent with well-known Hawking's semiclassical analysis. Moreover, the value of the Immirzi parameter of loop quantum gravity for {higher-dimensional case and four-dimensional U(1) case are} also obtained.

\end{abstract}


\maketitle

\section{Introduction}
%Outline: 1.the motivation for explaining black hole entropy in loop quantm grivaty. 2.overview of black hole entropy calculation in LQG,especially the method the role of Barbero-Immirzi parameter,. 3. the current development and importance of higher dimensional LQG. 4. the purpose of this paper.

{Black holes (BHs) as predicted by general relativity (GR) supply splendid platforms} for both experimental and theoretical physics. Recently, the Event Horizon Telescope (EHT) observations of the shadows of M87*\cite{Akiyama:2019cqa} and Sgr A*\cite{Akiyama:2022} have supported the existence of BHs and unveiled some mysteries about them. Theoretically, as {a} simple and strong gravitation object, a BH is a practical research object for GR, modified gravities, and quantum theories of gravity, especially after its thermodynamics was established by Bekenstein and Hawking~\cite{Bekenstein:1972tm,Bardeen:1973gs} in the 1970s. Whereafter, the thermodynamics was extended to the quasilocally defined boundary of the BH, namely isolated horizon (IH)~\cite{Ashtekar:1999yj}. The IH {enables us} to describe the BH as a physical object using local geometry which is independent of the spacetime outside the horizon. The Bekenstein-Hawking formula of BH entropy~\cite{Hawking:1974sw,PhysRevD.7.2333} brings GR, quantum mechanics, and statistical mechanics together; accounting for the statistical mechanics origin of BH entropy then becomes a great challenge for quantum theories of gravity.

As a well-known candidate for the quantum theory of gravity, there are various attempts made in the framework of loop quantum gravity (LQG)~\cite{As04,Ro04,Th07,Ma07} to account for the BH entropy by applying the notion of IH, for example, from the aspects of boundary topology theories~\cite{Ashtekar:1997yu,Ashtekar:2000eq,Meissner:2004ju,Engle:2009vc,Engle:2010kt,Wang:2014oua}, state counting methods~\cite{FernandoBarbero:2009ai,Agullo:2008yv,BarberoG.:2008ue,Agullo:2010zz}, and symmetries of IH~\cite{Ashtekar:2004nd,Song:2020arr}. Usually, in previous examples, the leading orders of entropy will agree with the Bekenstein-Hawking entropy if and only if the Immirzi parameter $\gamma$ takes some special value $\gamma_0$~\cite{Ashtekar:2000eq,Meissner:2004ju,FernandoBarbero:2009ai,Agullo:2010zz}. It gives a way to fix the free parameter $\gamma$ in LQG. In Ref.~\cite{Perez11}, the authors proposed a new method for calculating the entropy by introducing the universal horizon temperature and the energy measured by a local observer. In this approach, the Bekenstein-Hawking entropy could be reached for arbitrary values of $\gamma$; $\gamma$ occurs in the semiclassical correction term of statistic entropy via chemical potential. Then, the value of the Immirzi parameter can be fixed, if one further requires the classical formula of entropy is exactly the Bekenstein-Hawking one. This requirement is reasonable because if chemical potential is nonvanishing, one could achieve a lower energy by adding or removing particles (named punctures or intersections in loop quantized black hole depending on the boundary theory used). Then, in four-dimensional LQG, this {SU(2) counting} method leads to the following equation determining the value of Barbero-Immirzi parameter as~\cite{Perez11}
\ba
\sum_{j}(2j+1)e^{-2\pi\gamma\sqrt{j(j+1)}}=1
\ea with $j$ being half integers. This equation was also obtained in all previous {SU(2) }state counting of BH entropy in the LQG method \cite{Agullo:2008yv,Ghosh:2006ph}. {Except for the SU(2) Chern-Simons theory description of the quantum IH, the U(1) Chern-Simons description is also widely acknowledged~\cite{Ashtekar:1997yu,Meissner:2004ju,Domagala:2004jt}. Since the types of the constraints in U(1) and SU(2) descriptions are quite different, it is worth investigating whether the state counting in these two descriptions can be unified.}% through the method proposed in Ref.~\cite{Perez11}}

{Higher-dimensional gravity as an extension of general relativity has received wide attention. It is motivated from the Kaluza-Klein theory~\cite{appelquist1987modern} where apparently unrelated physical phenomena, namely gravity and Maxwell theory, can be unified in a higher-dimensional theory.} The loop quantization of higher-dimensional spacetime has been achieved in Refs.~\cite{Bodendorfer:2011nv,Bodendorfer:2011nw,Bodendorfer:2011nx,Bodendorfer:2011ny}. It is further applied to higher-dimensional cosmological models~\cite{Zhang16} and coherent states for semiclassical analysis \cite{Long:2019nkf,Long:2020euh,Long:2021xjm,Long:2021lmd,Long:2022cex}. Even though there are many achievements in higher-dimensional LQG, the first law and entropy of higher-dimensional BHs in the framework of LQG is still an open issue. It also leaves the Immirzi parameter {undetermined} in higher-dimensional LQG.

The aim of this article is to extend the IH {statistic mechanics} in LQG obtained in Ref.~\cite{Perez11} into higher-dimensional spacetime {and U(1) description in four-dimensional spacetime}, and fix the Immirzi parameter. This article is organized as follows: In Sec.~\ref{Section2}, we perform the local version of the first law measured by stationary observers in higher-dimensional spacetime. Utilizing the area operator in higher-dimensional LQG and the variational method, the quantum-corrected entropy and the first law are given in Sec.~\ref{Section3}. The value of the Immirzi parameter is analyzed in Sec.~\ref{Section4}. {We extend the statistic mechanics to the U(1) description in Sec.~\ref{Section5}} and conclude in Sec.~\ref{Section6}. Throughout the paper, we work in the unit of $c=k_B=1$.

\section{The classical configuration of black holes in higher dimensions}\label{Section2}

Let us consider the $(D+1)$-dimensional Schwarzschild spacetime. The metric expressed under spherically symmetric coordinates is~\cite{Myers:1986un,Horowitz12}
\be\label{metric}
ds^2=-(1-\big(\frac{r_H}{r}\big)^{D-2})\mathrm{d}t^2+(1-\big(\frac{r_H}{r}\big)^{D-2})^{-1}\mathrm{d}r^2+r^2\mathrm{d}\omega^2_{D-1},
\ee
where $r_H$ is the horizon radius and $\mathrm d\omega^2_{D-1}$%$=d\theta_1^2+\sin^2\theta_1 d\theta_2^2+\sin^2\theta_1\sin^2\theta_2d\theta_3^2+\cdots+\sin^2\theta_1\sin^2\theta_2\dots\sin^2\theta_{D-2}d\theta_{D-1}$,
is the line element of the unit $(D-1)$-sphere. Let $\Omega_{D-1}=2\pi^{D/2}/\Gamma(D/2)$, with $\Gamma(n)$ being the gamma function, denote the area of the unit $(D-1)$-sphere. Then,  the  BH horizon radius $r_H$ matches the BH mass $M$ through
\be\label{rad}
r_H=\left[\frac{16\pi G}{(D-1)\Omega_{D-1}}M\right]^{\frac{1}{D-2}},
\ee
where the $(D+1)$-dimensional gravitational constant in terms of the Plank length $\ell_P$ is $G =c^3\ell_P^{D-1}/\hbar$. For further convenience, we introduce the horizon area $A$, the surface gravity $\kappa_H$ and the Komar mass $E_H$. The relation between the horizon area and $r_H$ is $A=\Omega_{D-1}r_H^{D-1}$.  The surface gravity is defined by
\begin{equation}\label{eq:kappaH}
\kappa_H^2=-\frac{1}{2}\nabla_a \xi_b\nabla^a\xi^b
\end{equation}
where $\xi=\partial_t$ is the Killing vector which is timelike outside the horizon and normalized at the null infinity. Then,
given the metric \eqref{metric}, one gets
\be
\kappa_H=\frac{D-2}{2}\frac{1}{r_H}.
\ee
For the Komar mass $E_H$, one has
\be\label{energy}
E_H\equiv-\frac{D-1}{16\pi G(D-2)}\int_S \nabla^a \xi^b\,\mathrm{d}S_{ab}=M,
\ee
where $S$ is an arbitrary $(D-1)$-sphere surrounding the BH and the area element $\mathrm{d}S_{ab}=t_a\wedge r_b$ with $t_a$ and $r_b$ being the unit conormal vectors to $S$.

The classical version of the first law of BH thermodynamics have been extended to higher dimensions~\cite{Myers:1986un,Lewandowski:2004sh,Korzynski:2004gr,Ashtekar:2006iw,Zhang:2021umq}. Omitting the work term, one obtains%\footnote{one could redefined the gravitational constant by $\tilde{G}=\frac{D-1}{2(D-2)}G$, such that the coefficient of the first law takes the usual form $1/8\pi \tilde{G}$. In this case, the integral form for the first law will have an additional coefficient containing $D$. More specifically, one couldn't eliminate $D$ in both the differential and the integral forms of the first law, through redefine the gravitational constant. In order that the Plank length and the reduced Plank constant have a simpler relation, when setting $G=1=c$, we choose the definition of $G =c^3\ell_P^{D-1}/\hbar$.}
\be\label{1st}
\mathrm{d} E_{H}=\frac{\kappa_H}{8\pi G }\mathrm{d} A.
\ee
%G =\frac{2\pi}{\Omega_{D-1}}\frac{c^3\ell_P^{D-1}}{\hbar},
{With this equation, we can use the scaling arguments proposed {in Ref.}~\cite{waldsmarr} to get the Smarr formula}
\be\label{smarr}
E_H=\frac{D-1}{D-2}\frac{\kappa_H A}{8\pi G }.
\ee
{The Schwarzschild BH evaporates due to the Hawking radiation}. The equilibrium could be achieved by imbedding the BH into the Hartle-Hawking vacuum with the Hawking temperature
\be\label{tem}
T_H=\frac{\hbar \kappa_H}{2\pi}
\ee
at null infinity, which leads to a time symmetric thermal bath of radiation~\cite{PhysRevD.28.2960,Mersini-Houghton:2014zka}.
Comparing the first law \eqref{1st} of BH with that in the thermodynamics i.e., $\mathrm{d}E=T \mathrm{d}S$ and replacing $T$ by $T_H$ ({the temperature of equilibrium state consisting of BH  and Hartle-Hawking vacuum}), one can get the entropy of the BH as
\be\label{entropy-c}
S=\frac{A}{4\ell_P^{D-1}},
\ee
where $\ell_P^{D-1}=G\hbar$.
It is nothing but the generalization of the BH entropy in four dimensions.
%It should be noted that since one could rescaling $\epsilon$ in the expression of $\ell$, therefore, $\ell$ is independent of the mass or the entropy, which is could be regarded as a universal parameter . As explained in Ref.~\cite{Frodden:2011eb}, due to the universal parameter $\ell$ and hence universal temperature $T_H$, we get the first law \eqref{1st}.
{For simplicity, the classical  configuration containing the  surface gravity, the Komar mass, etc, is introduced via the Killing horizon. However,
} those quantities can also be defined on the IH, since the IH is the {generalization} of the Killing horizon {without} requiring a global Killing vector. {Without global structure required for normalizing the Killing vector at the null infinity one can not choose a unique null norm on the IH.} That is to say, the Killing vector $\xi^a$ on the IH can only be fixed up to a constant $\alpha$. As a consequence, the surface gravity given by \eqref{eq:kappaH}  contains a free  constant $\alpha$. The {IH} mass can be defined using the Hamiltonian method as the generator of a preferred time-translation {thereon}~\cite{Ashtekar:2000hw}. In this {approach}, the horizon mass is a secondary quantity expressed {in terms of} the fundamental quantities, {including the area, angular momentum and charges}, defined intrinsically at the horizon. If we think  of the nonrotating Killing horizon as a specific IH,  then the IH mass can also be defined on it. It turns out that the IH mass equals {the} Komar mass {up to the constant} $\alpha$. Since $\alpha$ is an overall factor, we will drop it { in the expression of the first law even for the case of IHs}. {Indeed, }the first law of BH thermodynamics has been extended to {higher-dimensional IHs}~\cite{Lewandowski:2004sh,Korzynski:2004gr,Ashtekar:2006iw}. {It is worth noting that since the quantities on the IH are independent of the spacetime outside it, we are allowed to quantize the  the IH and the outside spacetime separately in the framework of LQG, which is an advantage of the IH.}

\section{The statistic mechanics of quantum IH in higher dimensions}\label{Section3}

We reviewed the classical description of thermodynamics of higher-dimensional IHs in the last section. Now we use the classical configuration as physical input to investigate the statistic mechanics of the higher-dimensional quantum IHs via the microcanonical ensemble.
%do the same thing in quantum level by LQG and canonical ensembles

{To begin with, we describe the bulk  with the $(D+1)$-dimensional LQG with the structure group SO($D+1$) which was established in~\cite{Bodendorfer:2011nv,Bodendorfer:2011nw,Bodendorfer:2011nx,Bodendorfer:2011ny}. The phase-space variables are SO($D+1$) connection $A_{aIK}$ and its conjugate momentum $\pi^{aIK}$, where $a,b=1,\cdots,D$ are the  tensor indices on a spatial slice $\Sigma$ and $I,K = 0,\cdots, D$ are the internal SO($D + 1$) indices. The reason for choosing SO($D+1$) instead of SO($D$) as the gauge group is to match the degrees of freedom of the connection and its conjugate momentum~\cite{Bodendorfer:2011nv}. In the $(D+1)$-dimensional LQG, in addition to the Gauss, spatial diffeomorphism and scalar constraints, the simplicity constraints are imposed to form a first-class system.} After the simplicity constraints are solved, the remaining degrees of freedom \cite{Freidel99} on each edge are depicted by the Hilbert spaces
labeled by a non-negative integer $J$. The dimension of the $J$-space reads
\ba
\mathrm{dim}(\pi_J)=\frac{(J+D-2)!(2J+D-1)}{J!(D-1)!}.
\ea
Here the {integer $J$ plays a role somewhat like} the half-integer $j$ {labeling the representation space of SU(2), the structure group of the four-dimensional LQG}.
In $(D+1)$-dimensional LQG, the discrete spectrum of its $(D-1)$-dimensional area operator reads \cite{Bodendorfer:2011nx}
\ba
\Delta_{D-1}=8\pi G\hbar\gamma\sum_{J}\sqrt{J(J+D-1)}.
\ea

{For the (nonrotating) IH, it is described by the SO($D+1$) Chern-Simons theory as shown in \cite{Bodendorfer:2013jba,Bodendorfer:2013sja}.
In this theory, a quantum state of the IH is a set of punctures on it where each of these  punctures carries a real number. Now, let us identify the boundary of the bulk with the IH. The boundary condition requires that spin network edges in the bulk intersect the IH at the punctures.  Thus, each puncture inherits an integer $J$ from the edge  intersect it. The IH area $A$ then is proportional to the summation of those real numbers required by the boundary condition. Let $s_J$ be the number of punctures with the integer $J$. A quantum configuration is given by the sequence $\{s_J\}$.} Applying the strategy given in \cite{Perez11}, one has the following constraint on $\{s_J\}$
\ba
C_1:&&\quad\sum_{J}\sqrt{J(J+D-1)}s_J=\frac{A}{8\pi\gamma \ell^{D-1}_{p}},\label{C1}\\
C_2:&&\quad\sum_{J}s_J=N,\label{C2},
\ea
where $N$ denotes the total number of the punctures.  Moreover, the number of states $d[\{s_J\}]$ associated with the configuration $\{s_J\}$ is
\ba\label{eq:dsj}
d[\{s_J\}]=(N)!\prod_{J}\frac{1}{s_J!}\left[\frac{(J+D-2)!(2J+D-1)}{J!(D-1)!}\right]^{s_J}.
\ea The configuration which maximizes the entropy
$\log(d[\{s_J\}])$ and subjects to the above two constraints is exactly what we are looking for. Thus
we have the variational equation\ba
\delta\log(d[\{s_J\}])-\lambda \delta C_1-\sigma\delta C_2=0
\ea where $\lambda$ and $\sigma$ are the two Lagrange multipliers. By using Stirling's approximation formula
\ba
\log N!\thickapprox N(\log N-1)
\ea we can obtain the dominant configuration
\ba
\frac{s_J}{N}=\frac{(J+D-2)!(2J+D-1)}{J!(D-1)!}e^{-\lambda\sqrt{J(J+D-1)}-\sigma}.\label{sJ}
\ea
Summing over all possible $J$, we get
\ba\label{sum}
e^{-\sigma}\sum_{J}\frac{(J+D-2)!(2J+D-1)}{J!(D-1)!}e^{-\lambda\sqrt{J(J+D-1)}}=1,
\ea
by which we can solve $\sigma$
\be\label{sigma}
\sigma=\log\big[\sum_{J}\frac{(J+D-2)!(2J+D-1)}{J!(D-1)!}e^{-\lambda\sqrt{J(J+D-1)}}\big].
\ee
Let $\bar{d}$ denote the number  of states associated with the dominant configuration. Then $\bar d$ can be  calculated with Eq. \eqref{eq:dsj} once we  replace $s_J$ therein by that given in Eq. \eqref{sJ}. Applying the   Stirling's approximation formula again, with a straightforward calculation,  we get the entropy approximated by
\be\label{entropy-q}
S=\log \bar{d}=\lambda\frac{A}{8\pi\gamma\ell_P^{D-1}}+\sigma N.
\ee
where $\sigma$ depends on $\lambda$ due to Eq. \eqref{sigma}.

The Lagrange multiplier $\lambda$ can be expressed as a function of the inverse temperature  $\beta=\left(\frac{\partial S}{\partial E_H}\right)_N$
once we differentiate  Eq. \eqref{entropy-q} and employ Eq. \eqref{1st}. Then, we get
\begin{equation}
\lambda=\gamma\hbar\kappa_H\beta.
\end{equation}
%Substituting Eqs.~\eqref{entropy-q} and \eqref{energy} into the inverse temperature $\beta=\left(\frac{\partial S}{\partial E}\right)_N$, the Lagrange multiplier $\lambda$ can be expressed as a function of $\beta$, that is, $\lambda=\frac{(D-2)\gamma\hbar}{2r_H}\beta$.
{As explained in the last paragraph in Sec.~\ref{Section2}, the surface gravity and hence the Hawking temperature can be defined on the horizon properly (The constant dilation of the temperature is usually fixed to 1 in the literature~\cite{Ashtekar:2000eq}). Note that the system we consider is the horizon itself. Thus we can use the Hawking temperature as the system temperature.} Setting temperature $T_H$ as in Eq.~\eqref{tem}, one could get the Lagrange multipliers $\lambda$ as $\lambda=2\pi\gamma$. Substituting the value of $\lambda$ into Eq.~\eqref{sigma}, $\sigma$  can thus be obtained. Now, in terms of the free parameter $\gamma$ in LQG, we can express  entropy as
\ba
&&S=\frac{A}{4\ell_P^{D-1}}+N\sigma(\gamma), \quad \text{with}\label{entropy-q2}\\
&&\sigma(\gamma)=\log\big[\sum_{J}\frac{(J+D-2)!(2J+D-1)}{J!(D-1)!}e^{-2\pi\gamma\sqrt{J(J+D-1)}}\big]\label{eq:sigmagamma}.
\ea
One could see {that $\sigma(\gamma)$ plays the role of the chemical potential $\mu$ by its definition}
\be
\mu=-T_H\left.\frac{\partial S}{\partial N}\right|_E=-\frac{\hbar \kappa_H}{2\pi}\sigma(\gamma).
\ee
The chemical potential depends on $\gamma$ through {the} Lagrange multiplier $\sigma$.  {As shown in \eqref{eq:sigmagamma}, $\gamma$ performs as an exponential decay power. Thus, the  larger  value of $\gamma$ implies the smaller value of $\sigma$.} Differentiating Eq. \eqref{entropy-q2} and replacing $\mathrm{d} A$ by $\mathrm{d}E_H$ via \eqref{1st}, we get the first law
 \ba\label{1st-q}
\mathrm{d} E_H=T_H \mathrm{d} S+\mu \mathrm{d}N.
\ea

 %Since $\mu=-T_H\sigma(\gamma)$, the chemical potential could be negative, vanish or positive as the value of $\gamma$ increases. In this sense, the correct first law of a quantum IH mechanics should be
Substituting energy \eqref{energy} into \eqref{entropy-q2}, one could get the entropy at the quantum level as $S=\frac{D-2}{D-1}\beta E+\sigma N$. The coefficient {$(D -2)/(D-1)$} is caused by the dependence of the temperature on the horizon mass, which {matches} the one in the classical formula \eqref{smarr}. {We get a different integral formula of the first law with the one in Ref.~\cite{Perez11} when $D+1=4$, since the universal temperature is employed in that article while we use the Hawking temperature.} %, which is consisted with the one in 3+1 dimensional case in Ref.~\cite{Perez11} and is satisfied for dimensions $D+1\geqslant 4$.
{Compared with the classical  one \eqref{entropy-c}, the quantum entropy \eqref{entropy-q2} contains the extra term $N\sigma$. This term is interpreted as a quantum hair as in Ref.~\cite{Perez11}.
 }  This interpretation is convincing since $N$ is the total number of the punctures in the quantum description of the IH {and thus has no classical correspondence}. {Up to} the term of the quantum hair, the entropy \eqref{entropy-q2}  is consistent with the semiclassical Bekenstein-Hawking formula.

{For large BHs, the chemical potential $\mu$ and, thus, $\sigma$ which is proportional to $\mu$  should approach $0$ as the classical limit to achieve equilibrium \cite{Perez11}. In other words, $\gamma$ takes the value such that $\sigma\to 0$ as $N\to \infty$. }%(??? Is it  true that the classical limit means $N\to \infty$ so that  the Stirling formula can be applied?)
Hence we obtain by Eq. \eqref{sigma}
\ba\label{eqofgamma}
\sum_{J}\frac{(J+D-2)!(2J+D-1)}{J!(D-1)!}e^{-2\pi\gamma\sqrt{J(J+D-1)}}=1\label{gamma}
\ea which determines the value of $\gamma$ in higher-dimensional LQG.

\section{analysis of Immirzi parameter in higher-dimensional loop quantum gravity}\label{Section4}
%, $S_{J+1}/S_J$
As the most crucial free parameter in LQG, the Immirzi parameter occurs in the spectrums of geometric operators and the {Spin Foam dynamics}. The Immirzi parameter canot be fixed by LQG individually. In the last section, we obtain the equality \eqref{eqofgamma} that the Immirzi parameter should satisfy by comparing the statistic entropy and the Bekenstein-Hawking entropy. {Solving this equality,} we {can now} analyze the value of the Immirzi parameter in higher dimensions.
\begin{figure}[h!tb]
\centering
\includegraphics[width=0.6\linewidth]{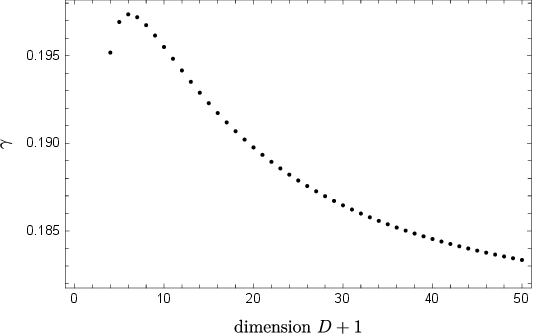}
\caption{The values of Immirzi parameter in $(D+1)$-dimensional spacetime.}
\label{fig:gamma}
\end{figure}
{In \figref{fig:gamma}, we show the  value of $\gamma$ for $4\leq D+1\leq 50$.} {As shown therein,} for $D=3$, {one gets} $\gamma=0.195177$. This result is different from the one obtained in {previous {papers} (see e.g.,~\cite{Meissner:2004ju,Domagala:2004jt,Perez11,Agullo:2008yv}), because we use a different gauge group in order to { adapt the higher-dimensional LQG}.  {Moreover, as shown in Fig.~\ref{fig:gamma},} the Immirzi parameter increases first and then decreases as the dimension increases. $\gamma$ takes the maximal value $\gamma_{\max}=0.197360$ {at $D+1=6$}.

\section{The statistic mechanics of quantum IH with U(1) gauge group}\label{Section5}

 There are two widely acknowledged choices of the gauge group of the Chern-Simons description of the quantum IH in four-dimensional spacetime, namely SU(2) and U(1). In Sec.~\ref{Section3}, we have extended the SU(2) case to SO($D+1$) to derive the entropy and the first law of the quantum IH in $D+1$ dimensions via the method proposed in Ref.~\cite{Perez11}. In this section, we will apply the same strategy to the U(1) case. In this case, the induced boundary degrees of freedom could be described by U(1) Chern-Simons theory at the classical level~\cite{Ashtekar:2000eq}.  {Due to} the boundary condition, counting the number of the U(1) Chern-Simons states converts to counting the number of the magnetic number sequences $(m_1, m_2,\cdots)$ under some constraints~\cite{Ashtekar:2000eq,Domagala:2004jt}, where $m_i$  {belongs {to}} $\{-j_i,-j_i+1,\cdots,j_i\}$ and  {is a} nonzero half integer. Denote the number of punctures  {carrying the} quantum  {magnetic number $k$} by $s_k$.
The constraints  {that the sequence $(m_1, m_2,\cdots)$ satisfies} are
\ba
C_1':&&\quad\sum_{k}\sqrt{|k|(|k|+1)}s_k\leqslant\frac{A}{8\pi\gamma \ell^{2}_{p}},\label{Cp1}\\
C_2':&&\quad\sum_{k}s_k=N, \label{Cp2}\\
C_3':&&\quad\sum_{k}k s_k=0. \label{Cp3},
\ea
where $C_1'$ and $C_2'$ are the analogues of $C_1$ and $C_2$ in  {the} SO($D+1$) case and $C_3'$ means that the sum of the angle deficits caused by the punctures vanishes modulo $4\pi$ in the U(1) case~\cite{Ashtekar:2000eq}.
 {Moreover, for the number $d[\{s_k\}]$ of the microstate with the configuration $\{s_k\}$}, one has
\be
d[\{s_k\}]=N!\prod_k\frac{2^{s_k}}{s_k!}. \label{dsk}
\ee

To find the configuration  which maximizes the value of $\log d[\{s_k\}]$ under the above constraints, one could still implement the variational method  {where the inequality constraint \eqref{Cp1} is dealt with}  by the Karush-Kuhn-Tucker (KKT) conditions \cite{karush1939minima,kuhn1951nonlinear}.  {More precisely, we} construct the Lagrangian function as
\be
L=\log d[\{s_k\}]-\lambda C_1'-\sigma C_2'-\omega C_3', \label{lagrangian}
\ee
 {with the Lagrangian multipliers $\lambda$, $\sigma$, and $\omega$. Then,}
according to the KKT conditions, the configuration we are looking for  {turns out to be} the solutions of constraints $C_1'$, $C_2'$ and $C_3'$ and the following equations
\ba
\log(2N) -\log s_k-\lambda\sqrt{|k|(|k|+1)}-\sigma-\omega k=0,\label{kkt-1}\\
\lambda\geqslant0,\label{kkt-2}\\
\lambda(\sum_{k}\sqrt{|k|(|k|+1)}s_k-\frac{A}{8\pi\gamma \ell^{2}_{p}})=0\label{kkt-3}.
\ea
There are two types of solutions: i) If the solution ${s_k}$ is  {in the interior of} the region restricted by $C_1'$, it is called the interior solution. In this case, the constraint Eq. \eqref{Cp1} is ineffective and $\lambda=0$. ii) If the solution is on the boundary of the region restricted by $C_1'$,  it called the boundary solution. In this case, Eq. \eqref{Cp1} is effective and $\lambda\neq0$.

Equation~\eqref{kkt-1} gives
\be
\frac{s_k}{N}=2e^{-\lambda\sqrt{|k|(|k|+1)}-\sigma-\omega k}.\label{sk1}
\ee
Substituting it into the constraint \eqref{Cp3}, we get
\be
\sum_k ke^{-\lambda\sqrt{|k|(|k|+1)}-\sigma-\omega k}= \sum_{|k|} |k|e^{-\lambda\sqrt{|k|(|k|+1)}-\sigma}(e^{-\omega|k|}-e^{\omega|k|})=0.
\ee
Since $|k|e^{-\lambda\sqrt{|k|(|k|+1)}-\sigma}>0$ for any nonzero $k$ and the sign of $e^{-\omega|k|}-e^{\omega|k|}$ is independent of $k$, the above equation holds if and only if $\omega=0$.  {Then Eq. \eqref{sk1} is simplified to}
\be
\frac{s_k}{N}=2e^{-\lambda\sqrt{|k|(|k|+1)}-\sigma}. \label{sk2}
\ee
Summing over all possible $k$, we get
\be
e^{-\sigma}\sum_k e^{-\lambda\sqrt{|k|(|k|+1)}}=\frac{1}{2}.
\ee

Now let us analyze which  {type the solution}  belongs to. If \eqref{sk2} is an interior solution, $\lambda$ must be zero. Then,
$s_k=2Ne^{-\sigma}$. Summing over all $k$, we get $\sum_k e^{-\sigma}=\frac{1}{2}$. Since there are  {infinitely many}  terms in this summation and $\sigma$ is a constant {independent of $k$}, one cannot find a solution for $\sigma$. Therefore, the solution \eqref{sk2} can only be  {a} boundary solution.
Substituting the solution \eqref{sk2} into  {Eq. \eqref{dsk}}, the approximated entropy is given as
\be
S= \log d[\{s_k\}]=\lambda\frac{A}{8\pi\gamma\ell_P^{2}}+\sigma N,
\ee
Setting the energy and the temperature as the Komar mass and the Hawking temperature {respectively}, we can get the entropy in the U(1) case as
\ba
&&S=\frac{A}{4\ell_P^{2}}+N\sigma(\gamma), \quad \text{with}\label{entropy-u1}\\
&&\sigma(\gamma)=\log\big[\sum_{k}2e^{-2\pi\gamma\sqrt{|k|(|k|+1)}}\big],
\ea
The first law  is $\mathrm{d} E_H=T_H\mathrm d S+\mu \mathrm dN$ with the  chemical potential $\mu=-T_H\sigma(\gamma)$. The entropy and the first law take the same formula as the ones in the SU(2) case {in regardless of} the expression of $\sigma(\gamma)$.

As explained above Eq.~\eqref{gamma}, the chemical potential must be zero classically. It determines the value of Barbero-Immirzi parameter through
\be
\sum_k e^{-2\pi\gamma\sqrt{|k|(|k|+1)}}=\frac{1}{2},
\ee
where $k\in \mathbb{N}/2$ and $k\neq0$. The approximated value of the Barbero-Immirzi parameter is $\gamma\approx 0.2375$.  This value matches the previous results in Refs.~\cite{Domagala:2004jt,Meissner:2004ju} by the U(1) state-counting method.

\section{Conclusion}\label{Section6}

%The thermodynamics of isolated horizon in higher dimension has been investigated via microcanonical ensemble and variational method in the framework of LQG.
The statistically mechanical calculation of the thermodynamical properties of nonrotating isolated horizons is studied in the framework of LQG. By employing the Hawking temperature and horizon mass of IHs, we establish the microcanonical ensemble associated with the system and extend the first law of quantum IH { with SU(2) gauge group in four-dimensional spacetime to the $(D+1)$-dimensional case with SO(D+1) group and four-dimensional case with U(1) group, that is, $\mathrm{d} E_H=T_H \mathrm{d} S+\mu \mathrm{d}N$}. It turns out that the higher-dimensional BH entropy and other thermodynamical quantities can be computed and consistent with well-known Hawking's semiclassical analysis. As a byproduct, the quantum hair of puncture $N$ has originated from the underlying quantum geometry, and hence, the first law of classical isolated horizons does not possess this term. Therefore, the only natural value of the chemical potential is zero at the classical level. This in turn fixes the value of the Immirzi parameter \eqref{eqofgamma} in higher-dimensional LQG.

Nonperturbative quantum theories of gravity with different values of $\gamma$ are not equivalent to each other, even though they could emerge the same classical theory. It means that physical input is needed to fix this free parameter. The role of the Immirzi parameter played in quantum gravity itself has already been a research topic \cite{Rovelli:2006,Dittrich:2012rj,Zhang162}, which is still understudying. In this paper, the Immirzi parameter affects the BH entropy through the quantum hair term, instead of the term proportional to the horizon area as in the earlier methods for calculating BH entropy in LQG did in four-dimensional cases~\cite{Ashtekar:2000eq,FernandoBarbero:2009ai} and higher-dimensional cases~\cite{Bodendorfer:2013sja,Wang:2014cga}. The primary reason for this difference is the various state counting method used in the literature.  Let us try to give a reasonable understanding of this new role. Since different quantum theories with different values of $\gamma$ should emerge the same classical theory, a naive perspective is that $\gamma$ should affect entropy as quantum correction so that it disappears in the classical limit. This is exactly the new role of $\gamma$ played in the  entropy \eqref{entropy-q2}, which is first proposed in Ref.~\cite{Perez11} and extended to higher-dimensional nonrotating IH in this article. In four-dimensional spacetime, different choices of gauge groups of boundary states, i.e., SO($D\!+\!1$=4), U(1) and SU(2), determine the different values of Barbero-Immirzi parameter through the spin distributions \eqref{sJ}, \eqref{sk2} and the one in Ref.~\cite{Perez11}. This means that different gauge groups lead to different spin distributions but give the same leading term of the number of microstates for a given horizon area. The differences among subleading terms in the cases of different gauge groups are worth further study.

%There are several previous works exploring the black hole entropy  in the higher-dimensional LQG.

This article only concerns the equilibrium state of BHs, we embed the BH into the Hartle-Hawking vacuum and keep its area and ,hence, mass fixed; therefore, black hole evaporation is not involved. LQG and its symmetric-reduced models give discrete BH mass~\cite{Zhang:2020qxw,Zhang:2021wex,Zhang:2021xoa,Zhang2020a}. It offers the possibility of realizing Hawking radiation in the framework of LQG.

\begin{acknowledgements}
This work is supported by National Natural Science Foundation of China (NSFC) through Grants No.12275087, No.11775082 and ``the Fundamental Research Funds for the Central Universities''. S. S. is also supported by NSFC Grand No. 12147167 and the project funded by China Postdoctoral Science Foundation Project No. 2021M700438. G. L. is supported by the project funded by China Postdoctoral Science Foundation through Grant No. 2021M691072, and the NSFC  with Grant No. 12047519. C. Z. also acknowledges the support by the Polish Narodowe Centrum Nauki, Grant No. 2018/30/Q/ST2/00811.

\end{acknowledgements}

%\bibliography{immizi}

\begin{thebibliography}{10}

\bibitem{Akiyama:2019cqa}
Kazunori Akiyama et~al.
\newblock {First M87 Event Horizon Telescope Results. I. The Shadow of the
  Supermassive Black Hole}.
\newblock {\em Astrophys. J.}, 875(1):L1, 2019.

\bibitem{Akiyama:2022}
Kazunori Akiyama et~al.
\newblock {First Sagittarius A* Event Horizon Telescope Results. I. The Shadow
  of the Supermassive Black Hole in the Center of the Milky Way}.
\newblock {\em Astrophys. J.}, 930(1):L12, 2022.

\bibitem{Bekenstein:1972tm}
J.~D. Bekenstein.
\newblock {Black holes and the second law}.
\newblock {\em Lett. Nuovo Cim.}, 4:737--740, 1972.

\bibitem{Bardeen:1973gs}
James~M. Bardeen, B.~Carter, and S.~W. Hawking.
\newblock {The Four laws of black hole mechanics}.
\newblock {\em Commun. Math. Phys.}, 31:161--170, 1973.

\bibitem{Ashtekar:1999yj}
Abhay Ashtekar, Christopher Beetle, and Stephen Fairhurst.
\newblock {Mechanics of isolated horizons}.
\newblock {\em Class. Quant. Grav.}, 17:253--298, 2000.

\bibitem{Hawking:1974sw}
S.~W. Hawking.
\newblock {Particle Creation by Black Holes}.
\newblock {\em Commun. Math. Phys.}, 43:199--220, 1975.
\newblock [,167(1975)].

\bibitem{PhysRevD.7.2333}
Jacob~D. Bekenstein.
\newblock Black holes and entropy.
\newblock {\em Phys. Rev. D}, 7:2333--2346, Apr 1973.

\bibitem{As04}
Abhay Ashtekar and Jerzy Lewandowski.
\newblock Background independent quantum gravity: a status report.
\newblock {\em Classical and Quantum Gravity}, 21(15):R53--R152, jul 2004.

\bibitem{Ro04}
Carlo Rovelli.
\newblock {\em quantum gravity}.
\newblock Cambridge University Press, 2005.

\bibitem{Th07}
Thomas Thiemann.
\newblock {\em Modern canonical quantum general relativity}.
\newblock Cambridge University Press, 2007.

\bibitem{Ma07}
Muxin Han, Yongge Ma, and Weiming Huang.
\newblock {Fundamental structure of loop quantum gravity}.
\newblock {\em Int. J. Mod. Phys. D}, 16:1397--1474, 2007.

\bibitem{Ashtekar:1997yu}
A.~Ashtekar, J.~Baez, A.~Corichi, and Kirill Krasnov.
\newblock {Quantum geometry and black hole entropy}.
\newblock {\em Phys. Rev. Lett.}, 80:904--907, 1998.

\bibitem{Ashtekar:2000eq}
A.~Ashtekar, John~C. Baez, and Kirill Krasnov.
\newblock {Quantum geometry of isolated horizons and black hole entropy}.
\newblock {\em Adv. Theor. Math. Phys.}, 4:1--94, 2000.

\bibitem{Meissner:2004ju}
Krzysztof~A. Meissner.
\newblock {Black hole entropy in loop quantum gravity}.
\newblock {\em Class. Quant. Grav.}, 21:5245--5252, 2004.

\bibitem{Engle:2009vc}
Jonathan Engle, Alejandro Perez, and Karim Noui.
\newblock {Black hole entropy and SU(2) Chern-Simons theory}.
\newblock {\em Phys. Rev. Lett.}, 105:031302, 2010.

\bibitem{Engle:2010kt}
Jonathan Engle, Karim Noui, Alejandro Perez, and Daniele Pranzetti.
\newblock {Black hole entropy from an SU(2)-invariant formulation of Type I
  isolated horizons}.
\newblock {\em Phys. Rev. D}, 82:044050, 2010.

\bibitem{Wang:2014oua}
Jingbo Wang, Yongge Ma, and Xu-An Zhao.
\newblock {BF theory explanation of the entropy for nonrotating isolated
  horizons}.
\newblock {\em Phys. Rev. D}, 89(8):084065, 2014.

\bibitem{FernandoBarbero:2009ai}
G.~J. Fernando~Barbero, Jerzy Lewandowski, and Eduardo J.~S. Villasenor.
\newblock {Flux-area operator and black hole entropy}.
\newblock {\em Phys. Rev. D}, 80:044016, 2009.

\bibitem{Agullo:2008yv}
Ivan Agullo, J.~Fernando Barbero~G., Jacobo Diaz-Polo, Enrique Fernandez-Borja,
  and Eduardo J.~S. Villasenor.
\newblock {Black hole state counting in LQG: A Number theoretical approach}.
\newblock {\em Phys. Rev. Lett.}, 100:211301, 2008.

\bibitem{BarberoG.:2008ue}
J.~Fernando Barbero~G. and Eduardo J.~S. Villasenor.
\newblock {Generating functions for black hole entropy in Loop Quantum
  Gravity}.
\newblock {\em Phys. Rev. D}, 77:121502, 2008.

\bibitem{Agullo:2010zz}
Ivan Agullo, J.~Fernando~Barbero, Enrique~F. Borja, Jacobo Diaz-Polo, and
  Eduardo J.~S. Villasenor.
\newblock {Detailed black hole state counting in loop quantum gravity}.
\newblock {\em Phys. Rev. D}, 82:084029, 2010.

\bibitem{Ashtekar:2004nd}
Abhay Ashtekar, Jonathan Engle, and Chris Van Den~Broeck.
\newblock {Quantum horizons and black hole entropy: Inclusion of distortion and
  rotation}.
\newblock {\em Class. Quant. Grav.}, 22:L27--L34, 2005.

\bibitem{Song:2020arr}
Shupeng Song, Haida Li, Yongge Ma, and Cong Zhang.
\newblock {Entropy of black holes with arbitrary shapes in loop quantum
  gravity}.
\newblock {\em Sci. China Phys. Mech. Astron.}, 64(12):120411, 2021.

\bibitem{Perez11}
Amit Ghosh and Alejandro Perez.
\newblock {Black hole entropy and isolated horizons thermodynamics}.
\newblock {\em Phys. Rev. Lett.}, 107:241301, 2011.
\newblock [Erratum: Phys.Rev.Lett. 108, 169901 (2012)].

\bibitem{Ghosh:2006ph}
A.~Ghosh and P.~Mitra.
\newblock {Counting black hole microscopic states in loop quantum gravity}.
\newblock {\em Phys. Rev. D}, 74:064026, 2006.

\bibitem{Domagala:2004jt}
Marcin Domagala and Jerzy Lewandowski.
\newblock {Black hole entropy from quantum geometry}.
\newblock {\em Class. Quant. Grav.}, 21:5233--5244, 2004.

\bibitem{appelquist1987modern}
Thomas Appelquist, Alan Chodos, and Peter George~Oliver Freund.
\newblock {\em Modern Kaluza-Klein Theories}, volume~65.
\newblock Addison-Wesley, 1987.

\bibitem{Bodendorfer:2011nv}
Norbert Bodendorfer, Thomas Thiemann, and Andreas Thurn.
\newblock {New Variables for Classical and Quantum Gravity in all Dimensions I.
  Hamiltonian Analysis}.
\newblock {\em Class. Quant. Grav.}, 30:045001, 2013.

\bibitem{Bodendorfer:2011nw}
Norbert Bodendorfer, Thomas Thiemann, and Andreas Thurn.
\newblock {New Variables for Classical and Quantum Gravity in all Dimensions
  II. Lagrangian Analysis}.
\newblock {\em Class. Quant. Grav.}, 30:045002, 2013.

\bibitem{Bodendorfer:2011nx}
Norbert Bodendorfer, Thomas Thiemann, and Andreas Thurn.
\newblock {New Variables for Classical and Quantum Gravity in all Dimensions
  III. Quantum Theory}.
\newblock {\em Class. Quant. Grav.}, 30:045003, 2013.

\bibitem{Bodendorfer:2011ny}
Norbert Bodendorfer, Thomas Thiemann, and Andreas Thurn.
\newblock {New Variables for Classical and Quantum Gravity in all Dimensions
  IV. Matter Coupling}.
\newblock {\em Class. Quant. Grav.}, 30:045004, 2013.

\bibitem{Zhang16}
Xiangdong Zhang.
\newblock {Higher dimensional Loop Quantum Cosmology}.
\newblock {\em Eur. Phys. J. C}, 76(7):395, 2016.

\bibitem{Long:2019nkf}
Gaoping Long, Chun-Yen Lin, and Yongge Ma.
\newblock {Coherent intertwiner solution of simplicity constraint in all
  dimensional loop quantum gravity}.
\newblock {\em Phys. Rev. D}, 100(6):064065, 2019.

\bibitem{Long:2020euh}
Gaoping Long and Norbert Bodendorfer.
\newblock {Perelomov-type coherent states of SO($D+1$) in all-dimensional loop
  quantum gravity}.
\newblock {\em Phys. Rev. D}, 102(12):126004, 2020.

\bibitem{Long:2021xjm}
Gaoping Long, Cong Zhang, and Xiangdong Zhang.
\newblock {Superposition type coherent states in all dimensional loop quantum
  gravity}.
\newblock {\em Phys. Rev. D}, 104(4):046014, 2021.

\bibitem{Long:2021lmd}
Gaoping Long, Xiangdong Zhang, and Cong Zhang.
\newblock {Twisted geometry coherent states in all dimensional loop quantum
  gravity: Construction and peakedness properties}.
\newblock {\em Phys. Rev. D}, 105(6):066021, 2022.

\bibitem{Long:2022cex}
Gaoping Long.
\newblock {Twisted geometry coherent states in all dimensional loop quantum
  gravity. II. Ehrenfest property}.
\newblock {\em Phys. Rev. D}, 106(6):066021, 2022.

\bibitem{Myers:1986un}
Robert~C. Myers and M.~J. Perry.
\newblock {Black Holes in Higher Dimensional Space-Times}.
\newblock {\em Annals Phys.}, 172:304, 1986.

\bibitem{Horowitz12}
Horowitz G.T., editor.
\newblock {\em Black holes in higher dimensions}.
\newblock Cambridge University Press, 2012.

\bibitem{Lewandowski:2004sh}
Jerzy Lewandowski and Tomasz Pawlowski.
\newblock {Quasi-local rotating black holes in higher dimension: Geometry}.
\newblock {\em Class. Quant. Grav.}, 22:1573--1598, 2005.

\bibitem{Korzynski:2004gr}
Mikolaj Korzynski, Jerzy Lewandowski, and Tomasz Pawlowski.
\newblock {Mechanics of multidimensional isolated horizons}.
\newblock {\em Class. Quant. Grav.}, 22:2001--2016, 2005.

\bibitem{Ashtekar:2006iw}
Abhay Ashtekar, Tomasz Pawlowski, and Chris Van Den~Broeck.
\newblock {Mechanics of higher-dimensional black holes in asymptotically
  anti-de Sitter space-times}.
\newblock {\em Class. Quant. Grav.}, 24:625--644, 2007.

\bibitem{Zhang:2021umq}
Xiangdong Zhang.
\newblock {Thermodynamics in new model of loop quantum cosmology}.
\newblock {\em Eur. Phys. J. C}, 81(2):117, 2021.

\bibitem{waldsmarr}
Daniel Sudarsky and Robert~M. Wald.
\newblock Mass formulas for stationary einstein-yang-mills black holes and a
  simple proof of two staticity theorems.
\newblock {\em Phys. Rev. D}, 47:R5209--R5213, Jun 1993.

\bibitem{PhysRevD.28.2960}
J.~B. Hartle and S.~W. Hawking.
\newblock Wave function of the universe.
\newblock {\em Phys. Rev. D}, 28:2960--2975, Dec 1983.

\bibitem{Mersini-Houghton:2014zka}
Laura Mersini-Houghton.
\newblock {Backreaction of Hawking Radiation on a Gravitationally Collapsing
  Star I: Black Holes?}
\newblock {\em Phys. Lett. B}, 738:61--67, 2014.

\bibitem{Ashtekar:2000hw}
Abhay Ashtekar, Stephen Fairhurst, and Badri Krishnan.
\newblock {Isolated horizons: Hamiltonian evolution and the first law}.
\newblock {\em Phys. Rev. D}, 62:104025, 2000.

\bibitem{Freidel99}
L~Freidel, K~Krasnov, and R~Puzio.
\newblock Bf description of higher-dimensional gravity theories.
\newblock {\em Advances in Theoretical and Mathematical Physics},
  3(5):1289--1324, 1999.

\bibitem{Bodendorfer:2013jba}
Norbert Bodendorfer, Thomas Thiemann, and Andreas Thurn.
\newblock {New Variables for Classical and Quantum Gravity in all Dimensions V.
  Isolated Horizon Boundary Degrees of Freedom}.
\newblock {\em Class. Quant. Grav.}, 31:055002, 2014.

\bibitem{Bodendorfer:2013sja}
Norbert Bodendorfer.
\newblock {Black hole entropy from loop quantum gravity in higher dimensions}.
\newblock {\em Phys. Lett. B}, 726:887--891, 2013.

\bibitem{karush1939minima}
William Karush.
\newblock Minima of functions of several variables with inequalities as side
  constraints.
\newblock {\em M. Sc. Dissertation. Dept. of Mathematics, Univ. of Chicago},
  1939.

\bibitem{kuhn1951nonlinear}
HW~Kuhn and AW~Tucker.
\newblock Nonlinear programming.
\newblock In {\em Proceedings of the Second Berkeley Symposium on Mathematical
  Statistics and Probability}, volume~2, pages 481--493. University of
  California Press, 1951.

\bibitem{Rovelli:2006}
Alejandro Perez and Carlo Rovelli.
\newblock Physical effects of the immirzi parameter in loop quantum gravity.
\newblock {\em Phys. Rev. D}, 73(4):044013 2006.

\bibitem{Dittrich:2012rj}
Bianca Dittrich and James~P. Ryan.
\newblock {On the role of the Barbero-Immirzi parameter in discrete quantum
  gravity}.
\newblock {\em Class. Quant. Grav.}, 30:095015, 2013.

\bibitem{Zhang162}
Xiangdong Zhang.
\newblock {Immirzi parameter and quasinormal modes in four and higher spacetime
  dimensions}.
\newblock {\em Front. Phys.}, 11(4):110401, 2016.

\bibitem{Wang:2014cga}
Jingbo Wang and Chao-Guang Huang.
\newblock {Entropy of higher dimensional nonrotating isolated horizons from
  loop quantum gravity}.
\newblock {\em Class. Quant. Grav.}, 32(3):035026, 2015.

\bibitem{Zhang:2020qxw}
Cong Zhang, Yongge Ma, Shupeng Song, and Xiangdong Zhang.
\newblock {Loop quantum Schwarzschild interior and black hole remnant}.
\newblock {\em Phys. Rev. D}, 102(4):041502, 2020.

\bibitem{Zhang:2021wex}
Cong Zhang, Yongge Ma, Shupeng Song, and Xiangdong Zhang.
\newblock {Loop quantum deparametrized Schwarzschild interior and discrete
  black hole mass}.
\newblock {\em Phys. Rev. D}, 105(2):024069, 2022.

\bibitem{Zhang:2021xoa}
Cong Zhang.
\newblock {Reduced phase space quantization of black holes: Path integrals and
  effective dynamics}.
\newblock {\em Phys. Rev. D}, 104(12):126003, 2021.

\bibitem{Zhang2020a}C. Zhang, X. Zhang,
\newblock { Quantum geometry and effective dynamics of Janis-Newman-Winicour singularities}.
\newblock {\em Phys. Rev. D}, 101, 086002, 2020.






\end{thebibliography}
%\bibliographystyle{unsrt}

\end{document}